\documentclass[11pt,letterpaper,notoc,nohyper]{JHEP3}
\usepackage{cite}
\usepackage{colordvi}
\usepackage{epsfig,bbm}
\input epsf.tex
\usepackage{graphics} 
\usepackage{amsmath}
\usepackage{latexsym}

\usepackage{graphicx}
\def\be{\begin{equation}}
\def\ee{\end{equation}}
\def\bea{\begin{eqnarray}}
\def\eea{\end{eqnarray}}
\newcommand{\eref}[1]{(\ref{#1})}

\newcommand\eqn[1]{\be\label{eq:#1}}
\newcommand\refeq[1]{eq. (\ref{eq:#1})}

\newcommand{\gsim}{ \mathop{}_{\textstyle \sim}^{\textstyle >} }
\newcommand{\lsim}{ \mathop{}_{\textstyle \sim}^{\textstyle <} }

\title{New   Regions for a Chameleon to Hide}
\author{
Baruch Feldman, Ann E. Nelson\\
Department of
  Physics, Box 1560, University of Washington, Seattle, WA 98195-1560
  \\  email:    \email{baruchf@u.washington.edu}, \email{anelson@phys.washington.edu} 
  } 
 \preprint{} 
 \abstract{We show that inclusion of an extremely small quartic coupling constant  in the potential for a nearly massless scalar field  greatly increases the experimentally allowed region for the  mass term and the coupling of the  field to matter.}

\begin{document}

\bibstyle{prstyle}

 \section{Introduction}
 \label{intro}
A large number of speculative theoretical ideas suggest the existence of new scalar fields, as reviewed in \cite{Moody:1984ba,Carroll:1998zi,Adelberger:2003zx}. A sufficiently light scalar will produce a new macroscopic force.  Such forces may lead to violations of    the   equivalence principle (EP) and the inverse square law (ISL).  For recent discussions of experimental searches for violation of the EP  see ref.\cite{Fischbach:1999bc,willbook,Fayet:2001nu,Will:2001mx,Will:2005yc}, while  recent reviews of searches for  ISL violation may be found in \cite{Fischbach:1999bc,Long:2003ta,Adelberger:2003zx}. 

Recently, Khoury and Weltman \cite{Khoury:2003aq} and Gubser and Khoury \cite{Gubser:2004uf} have shown that inclusion of non quadratic terms in the potential for a scalar field can  greatly alter the experimental constraints on its coupling and mass. In particular, the range and effective strength of such a ``chameleon'' force can drastically depend on the environment. Such effects have been previously explored in theories of time varying alpha   \cite{Mota:2003tc,Mota:2003tm,Steinhardt:2003iu}. Several subsequent works explored   such terms  in quintessence dark energy models \cite{Khoury:2003rn,Amendola:2003wa,Brax:2004qh,Wei:2004rw,Brax:2005ew}.
In this paper, we further explore the experimental implications of the chameleon effect. We consider the constraints on the couplings to ordinary matter of a nearly massless chameleon field with WEP preserving couplings.
Our results are dramatic: we find that the inclusion of an extremely tiny quartic coupling constant for the scalar field, as small as $10^{-53}$, weakens the constraints on the allowed coupling of the field to ordinary matter. We find  allowed parameter regions for new forces which have not been discovered previously. 
Similar conclusions would apply to EP constraints, although we do not do a numerical analysis of those constraints here.

\section{The Model}

We consider a real scalar field theory governed by the Lagrangian density 
\be
\label{eq:Lagrange}
-\mathcal{L} = \frac{1}{2} (\partial \phi)^2 +
V(\phi).  
\ee
We interpret $V(\phi)$ as the renormalized effective potential.  
We will consider the (time-independent) solutions of the equations of motion from this Lagrangian.  Our analysis uses classical field theory, which  is valid for a weakly coupled quantum  theory. 
We also assume that $\phi$ has been suitably shifted by a constant so $\phi=0$ is a global  minimum of $V$, and that  $V$ is analytic in the vicinity of $\phi=0$, which is equivalent to assuming that there is no degree of freedom which becomes massless at $\phi=0$.    

If we expand $V$ about $\phi=0$, the term linear in $\phi$ must vanish by minimization of $V$.  Furthermore,  we assume the minimum should be a global minimum, so that we do not have to consider whether the coupling to matter  could trigger formation of a catastrophic bubble causing tunneling to the true vacuum.  Therefore, the size of the coefficient of the cubic term is bounded in terms of the coefficients of the quartic and higher order terms.  
For arbitrarily weak fields produced by coupling $\phi$ to small source terms it suffices to consider only the quadratic term. Macroscopic sources produce larger field strengths for which the cubic and higher order terms also become important.  We will   stop  our expansion at quartic order. Because we have assumed $\phi=0$ is a global minimum of the potential, for a vacuum potential of the form 
\be V=\frac{1}{2} m^2 \phi^2 + \frac\mu3\phi^3+ \frac{\epsilon}{4} \phi^4\ee 
the coefficient of the cubic term satisfies
\be|\mu|\le3 \sqrt{m^2 \epsilon/2 \ .}\ee
The contribution of the  cubic term is then larger than that of the quadratic term only for field strengths large enough so that the quartic term is of at least comparable size.  Thus when computing the effects of the force due to $\phi$, provided the quadratic and quartic terms are included, 
  neglecting the   cubic terms   will lead to errors that are of most of order one. 
  
  We assume $\phi$  couples to the fields of the standard model. Inside most ordinary forms of macroscopic matter  the expectation value of any scalar operator involving the standard model gluon, quark  or electron fields is  approximately proportional to the mass density.
To estimate  at leading order  the modification of the constraints on a scalar field due to the inclusion of nonquadratic terms in the potential it therefore suffices      to consider the effective potential
\eqn{def-V}
V(\phi) \equiv  \frac{1}{2} m^2 \phi^2 + \frac{\epsilon}{4} \phi^4 - \beta \phi \rho,
\ee
where $\rho$ is the mass density of matter.  The equation of motion for $\phi$ is then  
\eqn{diffeq}
\partial^2 \phi = m^2 \phi + \epsilon \phi^3 - \beta
\rho = \frac{dV}{d\phi}.
\ee
In the limit $\epsilon\rightarrow 0$ we recover the Yukawa theory which is generally  the form assumed for new forces. {\it However  in general  for a scalar field there is no  principle requiring that $\epsilon=0$}. In fact,  a nonzero $\epsilon$ is required to renormalize the theory.

It follows from the Lagrangian density (\ref{eq:Lagrange}) that a test mass $m'$ would experience a fifth force 
\be
\vec{F_\phi} = - \beta m' \vec{\nabla}  \phi .
\ee

Experimental tests of the gravitational inverse square law generally put constraints on the parameters $\alpha$ and $\lambda$ of an additional Yukawa term
\be
\label{eq:yukawa}
\vec{F} = G_N m_1 m_2 \vec{\nabla} \frac{\alpha e^{-r/\lambda}}{r} .
\ee
These parameters are   related to those of  \refeq{diffeq} by
\be
m = 1/\lambda,   = \sqrt{4 \pi G_N \alpha}
\ee
in the case $\epsilon=0$.

A Yukawa force falls off exponentially fast at distances larger than   $\lambda\equiv 1/m$.
The    $ {\epsilon}  \phi^3$ in the equation of motion  will cause $\phi$ to fall even faster, at least as fast as a Yukawa field with a mass of    $m_{\rm eff} =\sqrt{m^2 +\epsilon \phi^2}$.   Following ref. \cite{Khoury:2003aq}, we refer to  $\phi$  as a {\it chameleon} field because the associated force  has an effective range and strength that varies according to the distribution of matter.  
In the next sections we show that the constraints on $m$ and $\beta$ are generally  weakened by nonzero $\epsilon$, and present both analytic estimates and numerical computations of the effect.

\section{Analytical approximations}
\subsection{Inside a Large Object: `thin shell' approximation}

The effective range of the force between test bodies is set by the inverse square root of the curvature of the potential. The chameleon potential is higher order than quadratic, and so the curvature of the potential about a nonzero field value may be much larger than it is about zero. When the range becomes shorter than the size of the object sourcing the field, then  only a ``thin shell'' of the object \cite{Khoury:2003aq} of thickness equal to the effective range of the force acts as a source for the field.

For sufficiently large $\epsilon$,   we may use  a thin-shell approximation  to find the value of $\phi$ inside a large object whose density varies slowly relative to the effective range of the force.  As a result, $\phi$ inside the object varies slowly and remains always near the value $\phi_{max}$ that maximizes 
\be
\label{def_phimax}
-V(\phi) = -\frac{1}{2} m^2 \phi^2 - \frac{\epsilon}{4} \phi^4 + \beta \phi \rho.
\ee

One condition required for this this approximation to be valid is:
\be
\label{thinshell-curv}
m_{\rm eff} \equiv \sqrt{V''|_{\phi_{max}}} = \sqrt{m^2 + 3\epsilon\phi_{max}^2}  > \frac{1}{\ell} 
\ee
where $\ell$ is the  scale over which the source density varies 
$$1/\ell\equiv {\rho'\over\rho}\ .$$  Note $\ell< R$, the radius of the source.
The left hand side of (\ref{thinshell-curv}) is an effective mass for the $\phi$ field, which depends on the density $\rho$ of matter through the maximization of eq. (\ref{def_phimax}).      

This approximation to minimizing the total energy  neglects the energy of the gradient terms in the field. Since the field must go to zero far from the source, the gradient energy outside the object may not be negligible. A second condition for the validity of the approximation is
\be
\label{thinshell-mag}
\phi_{max} < \phi_{Y}(R).
\ee
with $\phi_Y(R) \approx \frac{\beta M e^{-m R}}{4 \pi R}$ the Yukawa potential at the surface, and $M$ is the total mass of the source.  This condition ensures that, as $\epsilon \rightarrow 0$, we recover the Yukawa case.  

Maximization of (\ref{def_phimax}) determines the overall strength of the field produced by a source with a thin shell.  When
$$
\epsilon  \gg \left(\frac{m^3}{\beta \rho}\right)^2, 
$$
the mass term in $V(\phi)$ is negligible, and $\phi_{max}$ is given by
\be
\label{eq:phimax}
\phi_{max} \approx \left(\frac{\beta \rho}{\epsilon}\right)^{1/3}.
\ee

Substitution of \refeq{phimax} into the thin shell conditions \eref{thinshell-curv} and \eref{thinshell-mag} shows that the earth has a thin shell for $\alpha \epsilon \gsim 10^{-65}$.  

\subsection{Near surface behavior of a  thin-shell object}

For a spherical source,  static solutions to (\ref{eq:diffeq}) obey the ODE
\be
\label{eq:ODE}
\phi''(r) + \frac{2}{r}\phi'(r) = m^2 \phi + \epsilon \phi^3 - \beta \rho = \frac{dV}{d\phi}.
\ee
Exterior to a spherical object
we can get a good understanding of the solutions to \eref{eq:ODE} by considering the relative importance of the nonlinear term.  Since this term increases $m_{\rm eff}$, we expect $\phi$ to decay fast when \eref{eq:ODE} is nonlinear, and to approximate Yukawa behavior when the nonlinear term is unimportant.  Let us define 
\eqn{def-eta}
\eta^2 \equiv \frac{|\epsilon \phi^3|}{|m^2 \phi| + |\phi'/r|},
\ee
as a measure of the importance of nonlinearity in \eref{eq:ODE}. 

When the equation is sufficiently linear, the solution has Yukawa behavior, 
$\phi_{lin}' \approx -(m + 1/r) \phi_{lin}$.  For $r \ll 1/m$, we can self-consistently approximate 
\be
\label{eq:philin}
\phi_{lin} \sim \frac{\eta_{lin}}{\sqrt{\epsilon} r} ,
\ee
setting $\eta_{lin} \leq 1/5$.  If we attempted to apply this approximation right at the surface of a thin shell object, it would break down:
\be
\label{eq:surf-nonlin}
\eta_{surf} \sim \phi_{max} \sqrt{\epsilon} R  \sim \frac{m_{\rm eff} R }{\sqrt{3}} > \frac{m_{\rm eff} \ell }{\sqrt{3}} > \frac{1}{\sqrt{3}},  
\ee
by \eref{thinshell-curv}.  
So the conditions for a thin shell guarantee the nonlinear term is important near the surface.   

When the equation is sufficiently nonlinear, we can neglect the terms in the denominator of \refeq{def-eta}.  The LHS of \eref{eq:surf-nonlin} grows like $(\alpha \epsilon)^{1/6}$, so this is justified near earth's surface for most of our parameter space.  Then our differential equation is
$$
\phi'' \sim \epsilon \phi^3, 
$$
which would be exact for planar geometry and $m=0$.  This equation has the monotonically decaying or growing solutions
$$
\phi \sim \frac{1}{\pm \sqrt{\frac{\epsilon}{2}} r + C}.
$$
Matching to $\phi_{max}$ at $r=R $ gives 
\eqn{approx-sol}
\phi \sim \pm \frac{1}{\sqrt{\frac{\epsilon}{2}} (r - R  \pm \sqrt{\frac{2}{\epsilon}} \phi_{max}^{-1} )} 
.
\ee
So $\phi$ decays extemely fast, behaving like it has a pole $\sqrt{6}$ of a thin shell distance (i.e., $m_{\rm eff}^{-1}$) inside the surface.  
Eventually, $\eta$ falls to a value where the linear terms become important.  Neglecting $m$ and using \eref{eq:approx-sol} in \eref{eq:def-eta}, one finds this happens at
$$
r_{lin} \sim \frac{R  - \sqrt{6} / m_{\rm eff} }{1- 2/\eta^2},
$$ 
For a large thin shelled object,  $r_{lin}$ is of order $R$.  Even though $\phi_{max}$ increases with $\alpha\epsilon$, the shell thickness decreases, and $r_{lin}$ is insensitive to $\alpha\epsilon$.  For instance, as long as $\alpha \epsilon \gsim 10^{-52}$, $m_{\rm eff}$ of the earth is large enough that $r_{lin}$ is independent of $\alpha \epsilon$ to 1\%.  

Clearly, $\phi$ must fall fast as the shell gets thin.  One may wonder if this makes the chameleon force easy to detect near the earth's surface.  To answer this, we also use \refeq{approx-sol} to derive the field gradient near the surface:
$$
\phi'(R ) \sim - \frac{(3 \beta M )^{1/3}}{\epsilon^{1/6} (4 \pi)^{1/3} R ^2} \propto (\alpha/\epsilon)^{1/6}
$$
This shows that even though $\phi'(R_E)/\phi(R_E)$ grows with $\epsilon$, $\phi_{max}$ falls quickly enough so the surface force always decreases with $\epsilon$.  

The picture we have derived here is that $\phi$ decays very fast until the linear terms in the ODE become  important, then takes on Yukawa behavior.  The net effect of $\epsilon \neq 0$ weakens the effective coupling $\beta$.  All these results are confirmed by numerical simulations, although the location of linearization is not totally clear-cut.\footnote{  
It is difficult to describe analytically the region between $\eta\sim2$, where linear terms become important, and $\eta\sim1/5$, where they become the {\em only} important terms.  Numerical solution of the equations for $m=0$  shows that $\phi$ scales like $\phi_{lin} \propto \epsilon^{-1/2}$ instead of $\phi_{max} \propto \epsilon^{-1/3}$, for $r$ very near the surface.  But the $r$ for which $\phi$ scales with $\epsilon^{-1/2}$ is smaller than the value of $r$ for which  $\phi$ decreases with $r$ as $1/r$.  
}

\subsection{Effective coupling approximation}
\label{eff-coupl}
The results of the previous section show that
for a sufficiently massive thin shelled  object coupled to a chameleon field,    at a distance $r$ which is large compared to the radius of the object  but smaller than $1/m$, the field strength is  given by \be\phi_{lin}\approx \frac{c}{r\sqrt{\epsilon }} ,\ee where $c$ is a number of order 1. We can show that this result is always a good approximation for a spherical source whenever
\be\label{effective}  \epsilon \gsim \frac{1}{\alpha M^2 G_N}\ . \ee 
and $m$ is sufficiently small, and for any geometry at a distance which is much larger than the size of the source and shorter than $1/m$.    
Outside the source,  for $\epsilon=0$ the approximate solution to the field equations is  \be\phi_{Y}(r)=\beta \sqrt{G_N} M \frac{e^{-mr}}{r}\ .\ee 
For nonzero $\epsilon$,  $\phi$ will be smaller,  declining at least   as quickly as a Yukawa field with mass of order $m_{\rm eff}$.    Thus if $\phi$ is larger than $\phi_{lin}$,  $\phi$ will fall exponentially faster than $1/r$ until it decreases to a value of order $\phi_{lin}$.   

Note that  the differential equation \eref{eq:ODE} is covariant under the transformation 
$$
\phi\rightarrow k \phi, \epsilon\rightarrow \epsilon/k^2, \beta\rightarrow k \beta \ ,
$$  where $k$ is an arbitrary scaling factor,  which is consistent with the result that $\phi$ scales as $\epsilon^{-1/2}$.

 Comparing $\phi_{lin}$ with the Yukawa  field that would be produced for $\epsilon=0$ we see that in comparing the chameleon field strength with a Yukawa potential, at a distance which is large compared with the size of a source  one should use the {\it effective coupling}  
\be \beta_{\rm eff}\sim {\sqrt{4 \pi} \over M\sqrt{\epsilon} }.\ee
This approximation is good whenever the equations
\be
m\ll1/R
\ee
and \eref{effective} are satisfied.

Note that this effective coupling is independent of $\beta$, and are effects of a chameleon force are weaker than gravity for any object which is much more massive than the Planck mass divided by $\sqrt\epsilon$. Note also for any given object that the effective coupling depends only on $\epsilon$ and the mass of the object.

Thus   if we consider any given experimental test of the inverse square law  involving two sources  of mass $M_1$ and $M_2$ whose separation is larger than their size,  which bounds a certain $\alpha$ and $m$ for a Yukawa field, we conclude that for a chameleon field with the same $m$  we have a lower bound on $\epsilon$ which is
\be\label{constrainttwo}\epsilon > {1\over M_1 M_2 G_N  \alpha}\ .\ee 
Provided the inequality  \ref{constrainttwo} is  satisfied the chameleon is not constrained by the experiment in question. Similar conclusions would apply to equivalence principle tests, provided $\alpha$ is replaced by the product of $\alpha$ and the appropriate equivalence principle violating parameter.

\section{ Experimental Constraints}
This section derives the major constraints on chameleon forces from different experimental searches for fifth forces.  Again, these experiments generally put constraints on forces of the form \eref{eq:yukawa} for different length scales $\lambda = 1/m$.  Our calculations of the modified constraints for $\epsilon \neq 0$ are summarized in the figure.  

Typically, a Yukawa force is constrained only for $1/m$ of order the characteristic lengths in the experiment.  If $m$ is too large, a Yukawa force decays before it is detectable.  If $m$ is too small, the Yukawa force appears to obey  the inverse-square law.  Here we consider how $\epsilon \neq 0$ modifies this behavior (always weakening the force, as discussed above).  But we also consider how a modification to $G_N$ might be detected by affecting different  experimental measurements of $G_N$ differently.

\subsection{Lunar Laser Ranging}
\label{sec:LLR}

The Lunar Laser Ranging experiment \cite{Williams:1995nq} puts the tightest constraints of any experiment on a Yukawa force, and the fact that it is loosened for extremely small values of $\epsilon$ is our most prominent result.  For this reason, we chose $m \approx 1/R_{Earth-Moon}$, the most constrained value of $m$ when $\epsilon = 0$.  

The experiment carefully measures irregularities in the moon's orbit by reflecting a laser beam from a reflector on the moon.  This produces a constraint on Yukawa forces of $\alpha \lsim 10^{-11}$ for $m \approx 1/R_{Earth-Moon}$ \cite{Adelberger:2003zx}.  Let us apply the effective coupling approximation derived in section \ref{eff-coupl}.  We need to determine $\alpha_{\rm eff} $, the strength of the chameleon force on the earth-moon system relative to the strength of a Yukawa ($\epsilon=0$) force with $\alpha = 1$.  

We start by assuming both the earth ($\epsilon \gsim 10^{-65} / \alpha$) and moon ($\epsilon \gsim 10^{-61} / \alpha$) have thin shells.  The chameleon force is given by 
\be
\vec{F_C} = \vec{\nabla}_{R_{Earth-Moon}} \int d^3x \frac{1}{2} (\vec{\nabla} \phi)^2 + V(\phi),
\ee
where $\phi$ is a static solution to \refeq{diffeq} with both the earth and moon as sources.  As we saw above, a thin-shell source produces a field that falls to $\phi_{lin}$ in a distance roughly $\sim R_{source} / (1 - 2/\eta_{lin}^2)$ independent of $\epsilon$.  Farther from either source, the behavior is essentially linear, so $\phi \approx \phi_{Earth} + \phi_{Moon}$ with $\phi_{Earth}$ and $\phi_{Moon}$ single-source solutions, since the nonlinear regions do not overlap.  Since most of the overlap occurs in the linear region, one can by definition neglect the $\epsilon \phi^3$ term in $V(\phi)$, and the force too behaves linearly.  So it is separately proportional to $\beta_{\rm eff, i}$, the single source apparent strength of each field:
$$
\alpha_{\rm eff} = \frac{\beta_{\rm eff, E} \beta_{\rm eff, M}}{4 \pi G_N }.  
$$

After reaching $\phi_{lin}$, the single-source field has the Yukawa form, so 
$$
\beta_{\rm eff, i} = \beta \frac{\phi_{lin}}{\phi_Y(r_{lin})} \sim \frac{4 \pi \eta_{lin}}{\sqrt{\epsilon} M_i }, 
$$
where we have used $e^{-m r_{lin}} \sim 1$.  
Thus
 
\eqn{alpha-eff}
\alpha_{\rm eff} \approx \frac{4 \pi \eta_{lin}^2}{\epsilon G_N M_E M_M }
\ee
is the effective strength of the chameleon force (the $\epsilon \rightarrow 0$ limit clearly cannot be taken once a  thin shell approximation is used).  Our discussion above anticipates that $\alpha_{\rm eff}$ is insensitive to $\alpha$, since the large $r$ values of the field are insensitive to the surface value, which by \refeq{phimax} is $\phi_{max} \propto \alpha^{1/6}$.  
\footnote{A 10-fold rescaling of the surface value, equivalent to a rescaling of $\beta$ by $10^3$ or $\alpha$ by $10^6$, barely affects the later field.}  
Note, though, that the thin-shell conditions guarantee $\alpha_{\rm eff} < \alpha$.  Eq. (\ref{thinshell-mag}) in spherical geometry gives 
$$
\epsilon > \pi^3 \left(\frac{4}{\beta M}\right)^2,
$$
hence
$$
\alpha_{\rm eff} < \alpha \frac{\eta_{lin}^2}{\pi} \frac{\min(M_1, M_2)}{\max(M_1, M_2)} < \alpha/100 .
$$

Then the LLR constraint $\alpha_{\rm eff} < 10^{-11}$ gives:
\eqn{llr-constraint}
\epsilon > \frac{4 \pi 10^{11} \eta_{lin}^2}{G_N M_E M_M } \sim 5 \cdot 10^{-53}, \alpha\epsilon \gsim 10^{-61}.
\ee
when both sources have thin shells.  

For $\alpha \epsilon$ intermediate between the earth's and moon's thin shell values, we treat the moon as a point test mass.  Then 
$$
\alpha_{\rm eff} = \frac{ \beta \beta_{\rm eff, E}}{4 \pi G_N } = \sqrt{\frac{\alpha}{4 \pi G_N}} \beta_{\rm eff, E} < 10^{-11},
$$
so
\eqn{llr-constraint-2}
\alpha < \epsilon \frac{10^{44}}{(4\pi\eta)^2} \sim \epsilon (2 \cdot 10^{43}), 10^{-65} \lsim \alpha \epsilon \lsim 10^{-61}.
\ee
As  $\alpha \epsilon$ lowers to the value for which the earth acquires a thick shell, this constraint neatly approaches $10^{-11}$, the $\epsilon=0$ constraint.  

\subsection{Helioseismology}

One way to phrase the gravitational inverse-square law is to claim that the gravitational constant $G_N$ is indeed a constant.  So comparisons of measurements of $G_N$ with objects of greatly different size can constrain our model if the chameleon causes an apparent modification of $G_N$.  In section \ref{sec:lab-cav}, we will consider the constraints from comparing terrestrial Cavendish experiments, which involve source masses from $\sim10$ kg to $\sim10^4$ kg.  But first, we compare this whole group of experiments with $G_N$ determined by data from the sun, which has a mass $\sim10^{30}$ kg.  

The standard solar model balances pressure gradients against gravitation,  fixing the gravitational force between volume elements when the pressure is known.  Solar neutrino data constrains the temperature, and thus the gas pressure, to about 1\% \cite{Bahcall:2004qv}.  This constrains $G_N$, and not just the product $M G_N$, to 1\%, in agreement with lab-based Cavendish experiments.  

>From eqs. \eref{thinshell-mag} and \eref{thinshell-curv}, we find that the sun has a thin shell for $\alpha \epsilon \gsim 10^{-75}$.  A thin shell means that $\phi$ is constant inside the source (or varies quasi-statically for slowly changing density), so the chameleon force is then negligible in the sun's interior. (Strictly speaking, for thin shell we should compare the effective mass to the length scale $\ell$ for density variations, but we approximate by comparing to the solar radius).  The largest-scale Cavendish experiment has a thin shell for $\alpha \epsilon \gsim 10^{-23}$.  

In this and the following sections, we derive strict constraints and simplify our analyses by assuming the chameleon force completely negligible in each experiment when the source has a thin shell, and completely Yukawa when thin-shell fails.    This simple picture is valid except for $\alpha \epsilon$ near the endpoints of the thin shell regime, when $\eta \approx 1$ and the nonlinear and linear terms in the ODE are of the same order.  Improving the approximation would weaken the constraints in this regime, smoothing out their sharp edges.  

In this regime the Cavendish experiments see a $G_N$ modification of relative strength $\alpha$, while the sun sees only gravity.  Thus, we derive the constraint:
\eqn{helio-constraint}
{\rm For} \quad 10^{-75} \lsim \alpha \epsilon \lsim 10^{-23}, \quad \alpha \lsim 10^{-2} \  .
\ee

\subsection{Lab Cavendish experiments}
\label{sec:lab-cav}

The modern Cavendish experiments 
\cite{bagley,luo,Cohen:1987fr,luther,michaelis,mohr,izmailov,walesch,fitzgerald,Gundlach:2000rk,schlamminger} involve a range of techniques, geometries, and source masses from $\sim10$ kg to $\sim10^4$ kg.  There is some disagreement among results, and it is amusing to note that the chief outlier \cite{luo} differs from the accepted value by the right sign to be explained by a chameleon.  However, to derive the most stringent constraints, we discount this outlier and assume the other experiments accurate to their stated uncertainties.  The experiment with the smallest source masses \cite{Gundlach:2000rk} used 4 stainless steel spheres of mass 8 kg, giving a thin shell for $\alpha \epsilon \gsim 10^{-17}$.  The largest source mass was \cite{schlamminger} a pair of mercury tanks 7550 kg each, with a thin shell for $\alpha \epsilon \gsim 10^{-23}$.  

Then $\alpha$ must be such to make these two experiments agree within their fractional uncertainties of $10^{-5}$: 
\eqn{lab-constraints}
{\rm For} \quad 10^{-23} \lsim \alpha \epsilon \lsim 10^{-17}, \quad \alpha \lsim 10^{-5} \  .
\ee

\subsection{Ocean and Lake Experiments}

Zumberge et al \cite{Zumberge:1991sw} measured $g$ at varying depths in the ocean as a test of ISL and a determination of $G_N$.  Their value agrees with lab Cavendish experiments to about $3 \cdot 10^{-3}$.  Again, we expect no chameleon force in the ocean when the earth has a thin shell because $\phi$ is quasi-static in the interior.  This gives a constraint when the earth has a thin shell but laboratory experiments do not:
\eqn{ocean-constraint}
{\rm For} \quad 10^{-65} \lsim \alpha \epsilon \lsim 10^{-23}, \quad \alpha \lsim 3 \cdot 10^{-3} \  .
\ee

The lake experiments \cite{Fischbach:1999bc} use a lake as a source and typically find a value of $G_N$ that agrees with lab values to about $10^{-3}$.    This constraint is essentially subsumed by the ocean experiment.

\subsection{Tower Gravity Experiments}

Several experiments \cite{Fischbach:1999bc} measured constraints over scales of a few hundred meters by detecting deviations from an inverse-square falloff of $g$ as one ascends a tower.  
This is the case, near the surface of the earth, where the chameleon is strongest and falling fastest, among all the experiments.  

They observed limits on deviations of $\sim 10^{-7} g = 7 \cdot 10^{-17} / R_E$, setting $\hbar = c = 1$.  For the earth with a thin shell, we use our approximation \refeq{approx-sol} to compare the acceleration field a distance $\sim 10^2$ m $\sim 10^{-4} R_E \equiv \chi R_E $ above the surface to the expected ISL scaling:
\begin{eqnarray}
\beta |\phi'((1 + \chi) R_E) - \frac{\phi'(R_E)}{(1+\chi)^2}| &\sim& \frac{\beta}{\sqrt{\epsilon/2} R_E^2} \left|\frac{1}{(\chi + \epsilon^{-1/6}\sqrt{2}(\frac{4\pi}{3\beta M_E})^{1/3})^2}  - \frac{1}{(1+\chi)^2 2 \epsilon^{-1/3}(\frac{4\pi}{3\beta M_E})^{2/3}} \right| \nonumber \\ &\lsim &\frac{7 \cdot 10^{-17}}{R_E} 
\end{eqnarray}

Substituting in numbers:
\eqn{tower-constraint}
\sqrt{\frac{\alpha}{\epsilon}} \left| \frac{1}{(10^{-4} + (\alpha \epsilon)^{-1/6}(2.2\cdot10^{-11}))^2}- \frac{1}{(\epsilon\alpha)^{-1/3}(5\cdot10^{-22})} \right| \lsim 5 \cdot 10^{24}
\ee
For $\alpha \epsilon \ll 10^{-40}$, the shell becomes much thicker than $\chi R_E$, and the constraint approaches 
$$
\alpha \left|\frac{1}{(\alpha \epsilon)^{1/6}} - \frac{1}{2.2 \cdot 10^{-11}}\right| \lsim 10^7,  
$$
which becomes irrelevant for $\alpha \epsilon \sim 10^{-64}$, just as thin shell breaks down.  
For $\alpha \epsilon \gg 10^{-40}$, the first fraction in \refeq{tower-constraint} can be neglected, and $\alpha \lsim 10^4\epsilon^{1/5}$.  For $\epsilon \gsim 10^{-21}$, the overall chameleon strength is small enough that the constraint weakens to $\alpha \lsim 1$.  

\subsection{Lab Inverse-Square Law: Spero et al}

Some of the tightest lab constraints on Yukawa forces in the centimeter length scale come from Spero et al. \cite{spero,Hoskins:1985tn}.  This null experiment involved placing a 20 g test mass about 1.2 cm from the central axis of an Fe cylinder of inner radius $r_{in} = 3$ cm and outer radius $r_{out} = 4$ cm.  The experiment used a cancellation mass to cancel edge effects, so the force should vanish inside the cylinder for inverse-square law.  Spero et al. observed a torque corresponding to an acceleration field less than about $3\cdot10^{-12}$ m/s$^2$.  

As usual, our field becomes linear for a thick shell, so constraints are only different from those on a Yukawa field when $\epsilon$ large enough to give a thin shell, but small enough that the force is still detectable at the test mass position.  The thin shell criterion (\ref{thinshell-curv}) gives 
$$
\sqrt{3} (\beta \rho)^{1/3} \epsilon^{1/6} >  \frac{1}{r_{out} - r_{in}}, 
$$
so the Yukawa constraints are modified when $\alpha \epsilon \gsim 1.5\cdot10^{-14}$.  Notice that a large value of $\epsilon$ is needed to reduce the shell distance to a centimeter.  

The cylindrical geometry changes $2\phi'/r$ in \refeq{ODE} to $\phi'/r$, but we neglect this term anyway in our usual approximation \eref{eq:approx-sol} for very nonlinear fields.\footnote{
The validity of this approximation should extend $\frac{r_{out} - r_{in}}{1 - \eta^2 / 2} \sim$ a few cm, or most of the way inside the cylinder.  Numerical results confirm the approximation, at least to order of magnitude, about 80\% of the way from the wall to the axis.  }
Then the acceleration field at $r = 1.2$ cm is
$$
\beta \phi'(2 r_{in} / 5) \sim \frac{\sqrt{\epsilon/2} \beta}{(\sqrt{\epsilon/2}r_{in}(1 - \frac{2}{5}) + (\frac{\epsilon}{\beta\rho})^{1/3})^2} \lsim a_{obs} \approx 1.1\cdot10^{-29} / r_{in}
$$
This gives the constraint
\eqn{spero-constraint}
\frac{\alpha}{(\epsilon^{1/6} + \alpha^{-1/6} 6.6 \cdot 10^{-3})^4} \lsim \epsilon^{1/3} (2.5 \cdot 10^{6}).
\ee
This constraint equation can be solved numerically for $\alpha$ given $\epsilon$ satisfying thin shell.  The constraint would weaken even further for $\epsilon \gsim 3 \cdot 10^{-10}$, as the test mass acquired a thin shell.  

Note that \refeq{spero-constraint} depends sensitively on the position of the test mass.  Changing the position of the test mass would change the factor 2/5 in the first equation above, which would introduce a factor changing the relative importance of the two terms in the denominator of \refeq{spero-constraint}.  As the position of the test mass approached the inner wall, the constraint would strengthen to $\alpha \lsim \epsilon^{1/5} (4 \cdot 10^{-2})$.  However, to our knowledge no such experiment has been performed.  

\subsection{Lab Inverse-Square Law: Hoyle et al}
\label{sec:Hoyle}

The University of Washington Eotwash group \cite{Hoyle:2004cw} put tight constraints on a Yukawa force in the 100 micron range. The technique was to measure the torque on a torsion balance pendulum produced between a pair of  discs with holes bored in them.  In this setup, the ``missing mass" of the holes is the attractor for any field obeying a {\em linear} differential equation.  But it is difficult to model such a problem for a nonlinear field.  

The basic premise of this null experiment is that the force between large parallel plates is independent of separation for Newtonian gravity.  The setup made use of three discs, the lower two of which (7.8 mm and 1.8 mm thick Cu) were at a constant separation and had the holes offset azimuthally by 18 degrees.  The upper disc (2.0 mm thick Al) was at a variable separation of 0.2 mm to 11 mm and acted as a torsion pendulum.  Because the two lower discs were azimuthally rotated, they largely canceled each other's Newtonian torques on the pendulum.  But for a short-range force, only the upper disc would exert torque on the pendulum.  So we get to the core of the problem by calculating the separation dependence of the force between parallel plates for the chameleon field.\footnote{Of course, the cancellation of torques was not exact at all separations, but there is more cancellation for linear than   for nonlinear fields.}

The chameleon force becomes short range when linearity fails, which happens roughly when the source acquires a thin shell.  The two lower discs were flush, so we can consider them as a single attractor with holes partway through.  Then the experiment measured torque caused by the differential force from the parts of the attractor closer and farther from the pendulum.  For $\alpha \epsilon \lsim 6 \cdot 10^{-10}$, neither pendulum has a thin shell, and for sufficiently small $m$ we expect no ISL deviation in the experiment.  For $\epsilon$ very large, we expect the force to be negligible.  For intermediate values of $\epsilon$, we derive the constraints by computing the ratio of the forces from the two attractor discs and comparing to the Newtonian value of approximately 1 within the experimental sensitivity of about 1\%.  

For this planar one dimensional problem, our ODE becomes
$$
\phi''(x) = \frac{dV}{d\phi}
$$
with the origin chosen halfway between the plates.  It follows from the differential equation that the quantity
\be
\label{eq:C}
C \equiv \phi'^2 - 2 V(\phi)
\ee
is independent of position, in analogy to conservation of energy in a 1-d mechanics problem.  The potential energy density in the field is given by 
$$
u = \phi'^2 / 2 + V(\phi) = C/2 + 2 V(\phi).  
$$
Then the 1-dimensional chameleon ``force" between plates is the derivative of total potential energy with respect to separation:
$$
F_C = -\frac{dU}{d(\Delta x)} = -\frac{d}{d(\Delta x)} \int_{-\Delta x/2}^{\Delta x / 2} u dx 
= - \frac{1}{2}(u(\Delta x/2) + u(-\Delta x / 2)) - \int_{-\Delta x/2}^{\Delta x / 2} \frac{\partial u}{\partial \Delta x} dx 
.
$$

So far, this is exact, but integrating \eref{eq:C} and inverting for $\phi$ is an impractical way to find $F_C$.  If we apply thin shell, $\phi(-\Delta x/2) = \phi_{max, Cu}$, $\phi(-\Delta x/2) = \phi_{max, Al}$, then $u(\pm \Delta x/2)$ depends on separation only through $C/2$.  We further assume that, for our parameters, $\phi$ does not change much from $\phi_{max}$.  Then it is valid to approximate $\phi$ by a series solution to second order:
$$
\phi \approx \phi_0 + \phi_1 x + \phi_2 x^2 = \phi_0 + \left(\frac{\phi_{max,Cu} - \phi_{max,Al}}{\Delta x}\right) x + \left(\frac{\epsilon}{2} \phi_0^3 + \frac{m^2}{2} \phi_0 \right)  x^2
$$
and match to the thin shell solution:
$$
\frac{\phi_{max,Cu} + \phi_{max,Al}}{2}  \sim \left(\frac{\beta}{\epsilon}\right)^{1/3} \left(\frac{\rho_{Cu}^{1/3} + \rho_{Al}^{1/3}}{2}\right) \approx \frac{\alpha^{1/6}}{\epsilon^{1/3} r_0} \approx \phi_0 + \left(\frac{\epsilon}{8} \phi_0^3 + \frac{m^2}{8} \phi_0\right) (\Delta x)^2,
$$
with $r_0 \equiv 0.1$ mm.
The last equality can be solved numerically for $\phi_0$ as a function of $\epsilon$ and $\alpha$.  Then 
$$
C = (\phi_1 + 2 \phi_2 x)^2 - 2 V(\phi_0 + \phi_1 x + \phi_2 x^2)
$$
gives us the constraints.  We compare to the Newtonian 1-force
$$
F_N = - \frac{\pi G_N}{2} (\rho_2 t_2 - \rho_1 t_1)^2,
$$
with $t_i$ the thickness of plate $i$, by comparing 
$$
\frac{F_C(\Delta x) + F_N}{F_C(\Delta x + t_{upper}) + F_N}, 
$$ 
with $t_{upper}=1.8$ mm, to 1.  

We expect from \eref{eq:C} the ODE will become linear when $V(\phi) \sim V(\phi_{max}) \ll C$.  This occurs for $\alpha \epsilon \ll 4 \cdot 10^{-4}$ for separations of 0.2 mm and $\alpha \epsilon \ll 3 \cdot 10^{-14}$ for separations of 10 mm.  This latter inequality gives us some idea of the constraints since a linearized field is likely to behave like the Newtonian case.

Indeed, numerical simulations show that $\alpha \lsim 7 \cdot 10^{-11} / \epsilon$ for $\epsilon \lsim 10^{-8}$.  For larger $\epsilon$, the chameleon force weakens and the constraints become less relevant.  For $\epsilon=10^{-6}$, we have $\alpha \lsim 0.1$, and for $\epsilon=10^{-4}$, $\alpha \lsim 1$.  Beyond this, the parabolic approximation breaks down, but it is safe to assume that $\alpha \gsim 1$ is allowed.  

Hoyle et al also did a calibration using metal spheres of radii 4 mm and 2.5 cm, separated by 14 cm.  For $\alpha \lsim 10^{-15} / \epsilon$, all the calibration spheres would have thick shells, so the measured force   is $1 + \alpha$ times the gravitational force for   the calibration spheres as well as  for the pendula. There is therefore no constraint on $\alpha$ in this regime.  
  For $10^{-15} \lsim \alpha \epsilon \lsim  6 \cdot 10^{-10}$, the larger spheres but not the pendula would have a thin shell.  The chameleon then contributes more strongly to the pendulum force than to the calibration force. Since the  precision is about 1\% and no deviation is observed then
  $\alpha$ must be less than $10^{-2}$ in this regime.

\subsection{Lab Inverse-Square Law: Hoskins et al}

Hoskins et al \cite{Hoskins:1985tn} put constraints on the ISL at the 10 cm length scale.  They compared the force on a torsion pendulum from a 7 kg ``far" mass at about 105 cm separation to that from a ``near" mass at 5 cm and found $\alpha \lsim 10^{-3}$.  However, their experiment does not appear to constrain a chameleon force.  Since both masses were copper, they would have radii about 5 cm and 1 cm respectively.  But because each mass was located at least $\sim 5$ radii from the torsion pendulum, one would expect that the field from each mass at the torsion pendulum would always be linearized, either because each source had a thick shell or because the field from a thin shell source had already decayed past $\phi_{lin}$.  Since we use small $m$, a linear $\phi$ field would look like Newtonian gravity in this experiment.  We have already considered modification to $G_N$ in Section \ref{sec:lab-cav}.  

\subsection{Planetary}

Planetary constraints \cite{Fischbach:1999bc} are negligible at distance scales large compared to $1/m$  even for $\epsilon=0$, and as we have noted, the force always weakens with $\epsilon$.  If we considered the $1/m\approx$AU case, the discussion would parallel that for LLR.  

\subsection{Free Fall}

It is interesting to consider that, in the regime where two objects both have thin shells, the chameleon force behaves like a Yukawa interaction with $\alpha_{\rm eff}$ independent of $\alpha$ (see Sections \ref{eff-coupl} and \ref{sec:LLR}).  This is true when the separation is larger than $r_{lin} \sim R_{source} / (1 - 2/\eta_{lin}^2)$.  If we consider the regime where satellites have a thin shell ($\alpha \epsilon \gsim 10^{-18}$ for a 10 kg metal sphere), then satellites at heights greater than $\sim 5 R_E$ would experience a fifth force with $\alpha_{\rm eff}$ given by \refeq{alpha-eff} dependent on mass, affecting their periodicity.  

From the discussion in Section \ref{sec:LLR}, $\alpha_{\rm eff}$ would be down from $\alpha$ by a factor $M_{sat} / M_E \sim 10^{-20}$ even for the minimal value of $\epsilon$  which gives  a thin shell, and would decrease with $\epsilon$.  So such a fifth force would be negligible.  \FIGURE[t]{ 
 \centerline{\epsfxsize=4.5 in \epsfbox{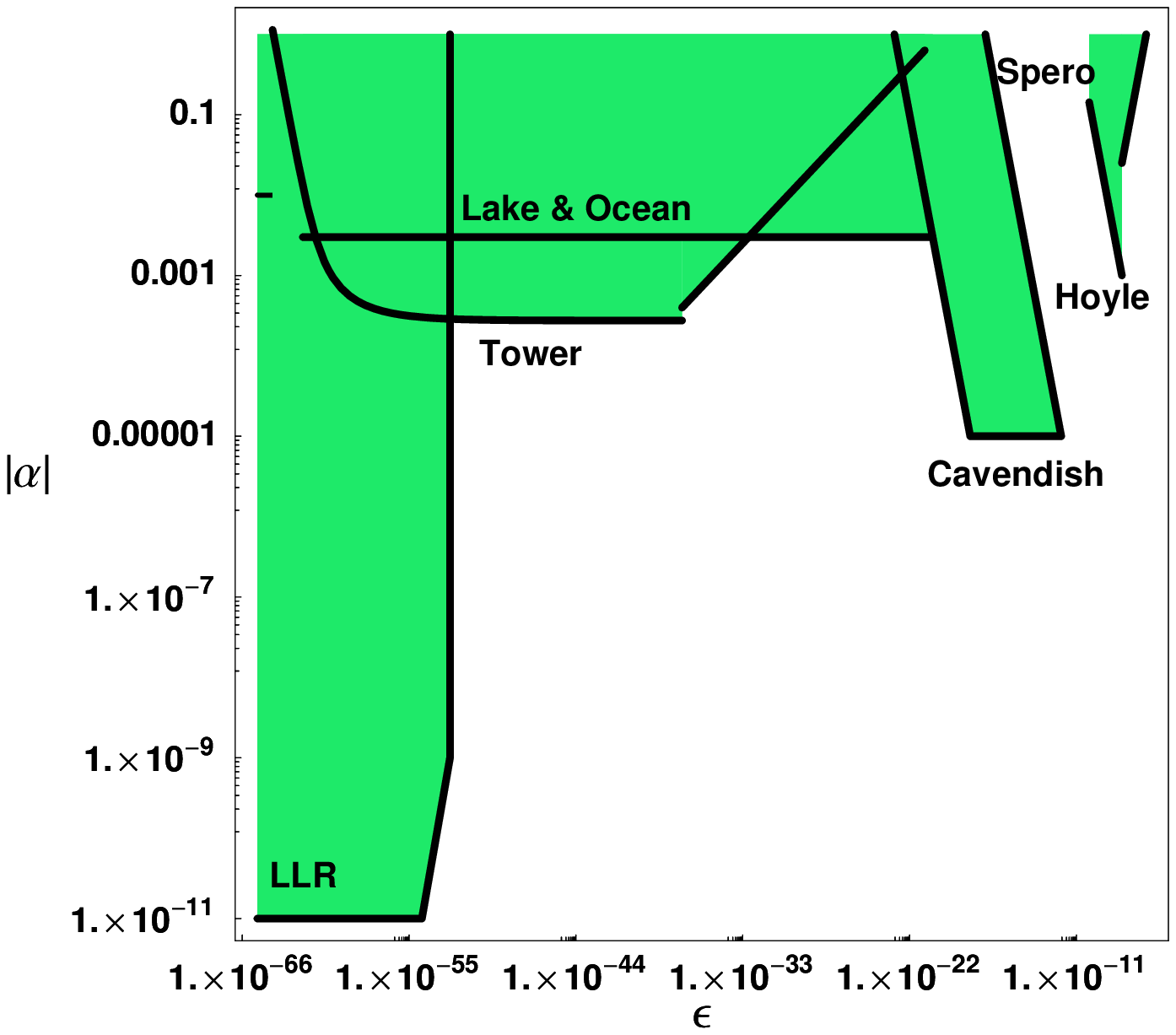}  } 
 \noindent\caption{Plot of allowed region for the coupling $\alpha$, as a function of the nonlinear term $\epsilon$, with the mass of the field set to the inverse earth-moon distance. The constraints do not change appreciably for massless or for somewhat more massive fields, except in the region $\epsilon < 10^{-50}.$ This entire region is forbidden by LLR when $\epsilon$=0.}
}
\section{Conclusion}
Currently, there is no evidence for deviations from Newtonian gravity, and this concordance is generally interpreted in terms of constraints on new forces. Usually, the   equations for any new force field are assumed to be linear. In this work, we have examined how such constraints would be reinterpreted  in the presence of a nonlinear, ``chameleon''  force. We note that some constraints on the strength of such a force become much weaker, even for extremely small nonlinear terms, and there is  room to hide a new long range force whose effects could be as large as   1\% of gravity between small objects.   

Although currently there is no compelling outlier, it is worth  emphasizing that for chameleon forces, the possibility   arises for different experiments to produce inconsistent results, when interpreted in terms of the Yukawa framework. Therefore outlying results on searches for new forces should not automatically be dismissed without further investigtion.
\section*{Acknowledgements}
We thank Eric Adelberger,  Jens Gundlach and Wick Haxton for useful discussions. 
This work was partially supported by the DOE under contract DE-FGO3-96-ER40956. A.N. would like to acknowledge the support of the Gugggenheim foundation.

\end{document}